\documentclass[twocolumn,showpacs,preprintnumbers,amsmath,amssymb,prb,aps]{revtex4-1}
\usepackage{epsfig}% Include figure files
\usepackage{epstopdf}
\usepackage{graphicx}% Include figure files
\usepackage{dcolumn}% Align table columns on decimal point
\usepackage{bm}% bold math
\usepackage{textcomp}
\usepackage{array}
\usepackage{subcaption}
\usepackage{booktabs}

\begin{document}

\title{TRACK: A python code for calculating the transport properties of correlated electron systems using Kubo formalism}

\author{Antik Sihi$^{1,}$}
\altaffiliation{sihiantik10@gmail.com}
\author{Sudhir K. Pandey$^{2,}$}
\altaffiliation{sudhir@iitmandi.ac.in}
\affiliation{$^{1}$School of Basic Sciences, Indian Institute of Technology Mandi, Kamand - 175075, India\\
$^{2}$School of Engineering, Indian Institute of Technology Mandi, Kamand - 175075, India}

\date{\today}
\small
\begin{abstract} 

  Exploring the transport properties of different materials brings new avenue for basic understanding of emergent phenomena and practical applications in many different fields. Here, we report a program named as TRACK (\textbf{TRA}nsport properties for \textbf{C}orrelated materials using \textbf{K}ubo formalism) which is written in Python 3 for calculating temperature dependent electrical conductivity, electronic part of thermal conductivity, Seebeck coefficient and Lorenz number. In this code, Kubo linear-response formalism is utilized for computing these parameters using both interacting and non-interacting electronic structure methods. The formula for transport coefficients is accordingly modified to obtain the transport parameters under relaxation time approximation using band-theory. The basic inputs of this program are the structural information, dense k-points sampling in the irreducible part of the Brillouin zone and the information of velocity matrix elements, which can be calculated using third-party \textit{ab-initio} package. TRACK is expected to calculate the transport properties of different class of materials. The code has been benchmarked by performing calculation on three different types  of materials namely Vanadium (V), FeSi and LaCoO$_3$, which are metal, semiconductor and Mott insulator, respectively. The temperature dependent behaviour of the transport coefficients for these materials show fairly good agreement with the corresponding experimental data.
  
\end{abstract}

\maketitle

\section{Introduction} 

  The world of materials is broadly divided into metal, semiconductor and insulator. Experimental study of different transport properties is one of the procedures to verify the properties of these classes of materials. In addition to this, when experiments of transport properties at low temperatures are performed, then many exotic phenomena like Kondo effect \cite{stewart_nfl,singh_nfl}, non-Fermi Liquid behaviour \cite{fisk_kondo}, superconductivity \cite{bednorz_sc} etc. reveal their existence. Temperature dependent experimental observation is necessary for understanding above mentioned properties. With the change in temperature the bands of quasiparticles (\textit{e.g.} electron, phonon, defects/impurity etc.) may alter their behaviour. The small changes in bands' behaviour will directly reflect on the transport properties of material. It happens because of the direct dependence of transport coefficients on band features. Therefore, proper estimation of temperature dependent band-structure of any material is important for theoretical calculation of transport properties. Along with the temperature effect, another bottleneck in theoretical prediction is the presence of different scattering sources in material. The intrinsic scattering mechanisms are electron-electron interaction (EEI), electron-phonon interaction (EPI) and phonon-phonon interaction (PPI). Further, electron-defects interaction (EDI) is also prominent as an extrinsic scattering source in the experimentally synthesized materials \cite{aschroft}. The EEI plays an important role in estimating the electronic structure of strongly and moderately correlated materials. Thus, investigation of transport properties will help us to understand the new emergent phenomena as well as the scattering mechanism involved in the material. 
  
  Nowadays, exploring the transport properties of materials is an important research area for basic understanding and application purposes.
Moreover, when the correlation effect and transport are merged then it opens a new area to understand the exotic phenomena as well as the materials through transport studies. For example, the Kondo metals show an upturn in resistivity ($\rho$) at low temperature, for non-Fermi liquid system $\rho$ changes with $T^\alpha$ ($\alpha\neq 2$) and also many strongly correlated materials show anomalous behaviour in transport properties \cite{fisk_kondo,imada_rev}. In present days, the applications of these materials have been explored in the field of thermoelectric (TE), optoelectronic etc \cite{coulter_opto,jtomczakRev}. The correlated thermoelectric materials (CTM) provide new avenue to overcome the limitations of transport properties of the existing uncorrelated semiconductors \cite{jtomczakRev}. In order to understand the transport behaviour, usually one needs to calculate electrical conductivity ($\sigma$), Seebeck coefficient ($S$) and thermal conductivity ($\kappa$). The $\kappa$ contains both electronic ($\kappa_e$) and phononic ($\kappa_{ph}$) contributions. To compute all these quantities except $\kappa_{ph}$, any \textit{ab-initio} electronic structure methodology is needed for obtaining the eigenstates of quasiparticle.
  
  In light of this, the density functional theory (DFT) is one of the widely used methodology, where the non-interacting single particle Kohn-Sham orbitals are considered as the eigenstates of the material \cite{payne_dft}. The semi-classical Boltzmann transport equations (BTE) is usually chosen to calculate the transport parameters by employing Kohn-Sham eigenstates. BoltzTraP \cite{boltztrap} and BOLTZWANN \cite{boltzwann} are most commonly used codes, which follow the above mentioned theory. Here, the effect of temperature is considered through the Fermi-Dirac distribution function. But in this scenario, the ground state electronic structure is used for the finite temperature calculation, which may not be desirable for the finite temperature transport properties calculation. In addition to this problem, the correlation effect within the partially filled $d$/$f$ electrons is not sufficiently considered in DFT. Therefore, it fails to provide proper electronic structure of strongly or moderately correlated electron systems. At this point, dynamical mean-field theory (DMFT) comes to tackle the Coulomb interactions of these correlated electrons using many-body interaction theory, where the other electrons are taken care by DFT in combined DFT+DMFT method \cite{kotliar_dmft,haule_dmft}. In such type of correlated materials, Kubo linear-response formalism (details of the methodology discussed in section II A) is usually employed for estimating the transport properties using the spectral function ($A{(\mathbf{k},\omega)}$) on real $\omega$ axis \cite{jtomczakRev}. The $A{(\mathbf{k},\omega)}$ can be obtained from any Green-function based electronic structure method like DFT+DMFT and $GW$ methods. Moreover, computing the transport coefficients using the self-energy in imaginary axis is possible, but this procedure needs lot of computational efforts for calculating the higher-order transport coefficients \cite{haule_dmft}. As per our knowledge, TRIQS/DFTTools is the only open-source code for correlated materials, where the Kubo formalism using $A{(\mathbf{k},\omega)}$ is utilized for calculating the different transport coefficients \cite{aichhorn_triqs}. This code is interfaced with WIEN2k for calculating the transport parameters \cite{wien2k}. In TRIQS/DFTTools code, the projected Wannier functions are implemented as the basis set for the correlated orbitals, which has followed the work of Aichhorn \textit{et al} \cite{aichhorn_imple}. However, Haule \textit{et al.} mentioned that this type of projection generally provides incorrect spectral weight but preserves the causal DMFT equations \cite{haule_dmft}. The eDMFTF code is developed by considering the new projection operator, which defines the connections between real-space objects and the corresponding orbital's counterpart. This new projection operator satisfies the causality of DMFT equations and leads to correct spectral weight. Thus, the eDMFTF code is the better choice to perform the electronic structure calculation for correlated materials \cite{haule_dmft}. In this case, the continuous-time quantum Monte Carlo (CTQMC) method is implemented as the impurity solver \cite{haule_ctqmc}. But to the best of our knowledge, this open-source code, which is interfaced with WIEN2k \cite{wien2k}, has no facility to compute the transport properties using Kubo formalism. Therefore, easily accessible open-source code is necessary for calculating the transport coefficient of correlated materials using the $A{(\mathbf{k},\omega)}$, which is obtained from eDMFTF code. The Kubo formalism is also possible to be used for any $GW$ based program, where $A{(\mathbf{k},\omega)}$ is computed.  
    
 Tomczak $et$ $al.$ showed that the Kubo formalism along with the consideration of vanishing scattering rate has capability to calculate the transport properties using the eigenstates of band-theory \cite{jtomczakRev}. In case of BoltzTraP \cite{boltztrap} and BOLTZWANN \cite{boltzwann} codes, different transport properties are calculated based on the semi-classical BTE using constant relaxation time approximation (CRTA). However, in general, the relaxation time ($\tau$) is function of direction (\textbf{r}) and energy ($\omega$). But, for homogeneous system, direction independent $\tau$ is usually considered \cite{aschroft}. Therefore, implementation of $\omega$ dependent $\tau$ is desirable for calculating the transport properties when eigenstates of DFT are used. In BoltzTraP2 code, the relaxation time approximation (RTA) along with CRTA is implemented for finding the electron-phonon scattering rate where $\tau$ is a function of density of states \cite{BoltzTraP2}. Moreover, the choice of basis function always plays important role in calculating the physical properties. The Kohn-Sham orbitals are required in BoltzTraP \cite{boltztrap} and BoltzTraP2 \cite{BoltzTraP2}, where the calculation to get Kohn-Sham orbitals for highly dense k-mesh is time consuming. To reduce the computational time, the maximally-localized Wannier function is utilized in BOLTZWANN \cite{boltzwann} for making the model Hamiltonian in low-energy region. However, proper choice of Wannier functions for the Kohn-Sham orbitals is a tedious task and the all electron eigenstates are always preferable than the eigenstates obtained from model Hamiltonian for calculating any physical properties. Using the Kubo formalism, the implementation of $\omega$ dependent $\tau$ is possible using Kohn-Sham orbitals, which brings to explore more insight of any material from the transport properties simply at level of DFT. As per our present knowledge, no open-source code is available, where the above mentioned technique is used to study the transport coefficients. 
 
  Here, a Python 3 code named as TRACK (formally called as \textbf{TRA}nsport properties for \textbf{C}orrelated materials using \textbf{K}ubo formalism), which is interfaced with WIEN2k \cite{wien2k} and eDMFTF \cite{haule_dmft} codes, is reported to study $S$, $\sigma/\tau$, $\kappa_e/\tau$ and Lorenz number ($L$), respectively. Kubo linear-response formalism is used for exploring the basic insights of transport parameters with considering non-interacting ($e.g.$ DFT) and interacting ($e.g.$ DFT+DMFT) electronic structure methods. In order to benchmark the code, we choose different type of non-magnetic correlated materials such as Vanadium (V), FeSi and LaCoO$_3$, which are prototype of well known correlated metal, semiconductor and insulator, respectively. The calculated values from our code are also compared with the results obtained from BoltzTraP \cite{boltztrap} calculations for all these materials.

\begin{figure}[]
   \begin{center}
   \begin{subfigure}{0.85\linewidth}
   \includegraphics[width=0.60\linewidth, height=1.8cm]{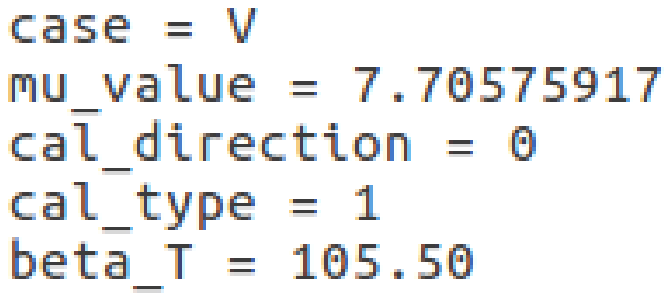}
   \caption{}
   \label{fig:} 
\end{subfigure}
\begin{subfigure}{0.90\linewidth}
  \begin{subfigure}{0.45\linewidth}
   \includegraphics[width=0.95\linewidth, height=4.2cm]{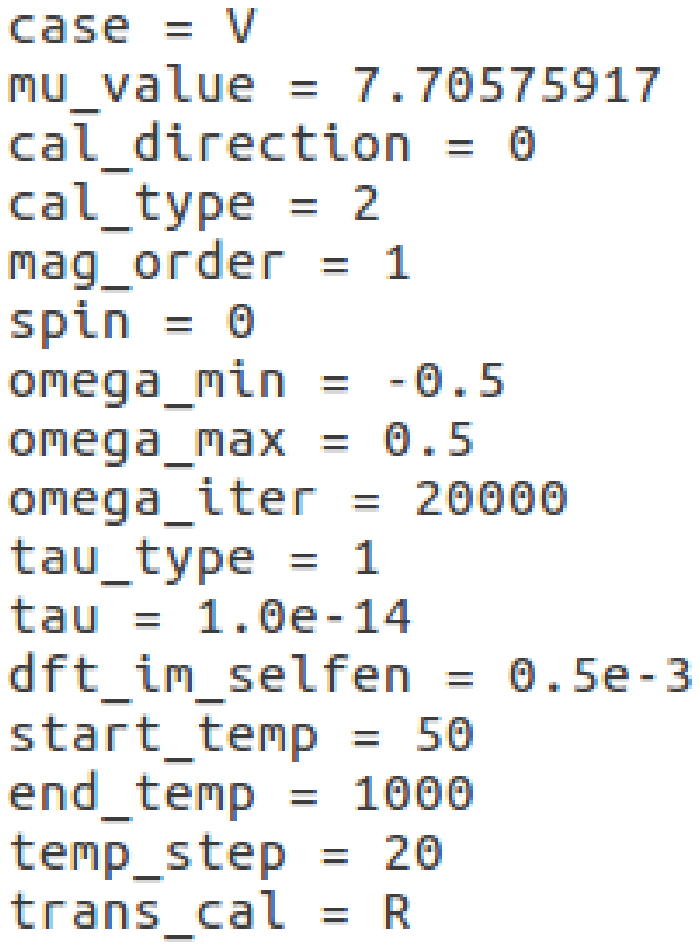}
   \caption{}
   \label{fig:}
   \end{subfigure}
   \begin{subfigure}{0.45\linewidth}
   \includegraphics[width=0.95\linewidth, height=4.2cm]{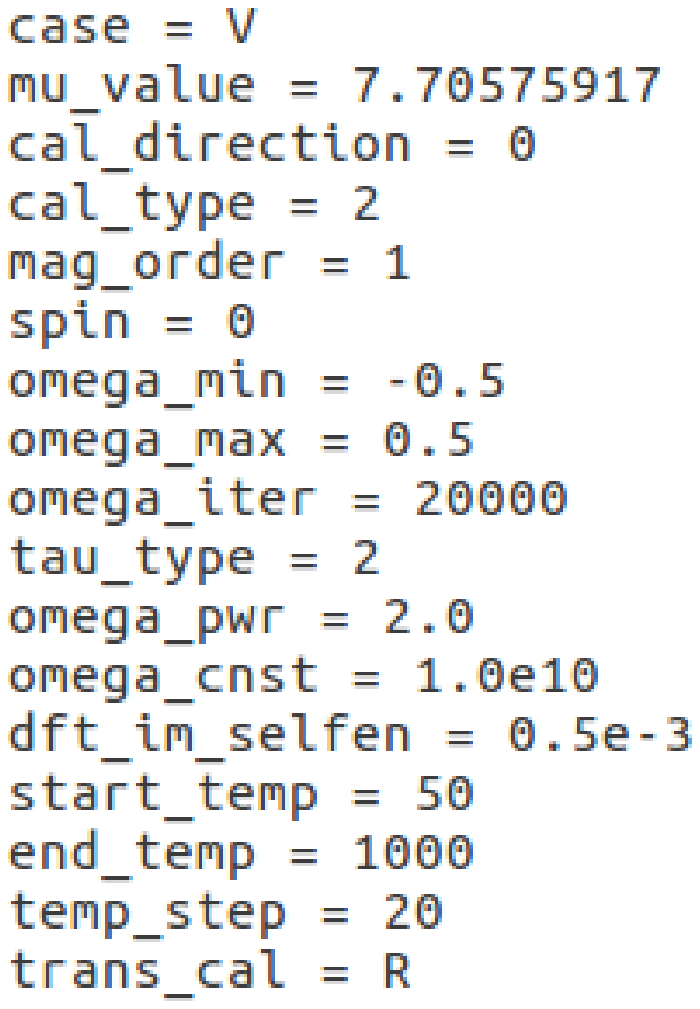}
   \caption{}
   \label{fig:}
   \end{subfigure}
\end{subfigure}
\caption{\small {Screenshot of input files when using the eigenvalues of (a) many-body theory and (b) \& (c) band-theory.}}
   \end{center}
\end{figure}

\begin{figure}
  \begin{center}
    \includegraphics[width=0.89\linewidth, height=9.0cm]{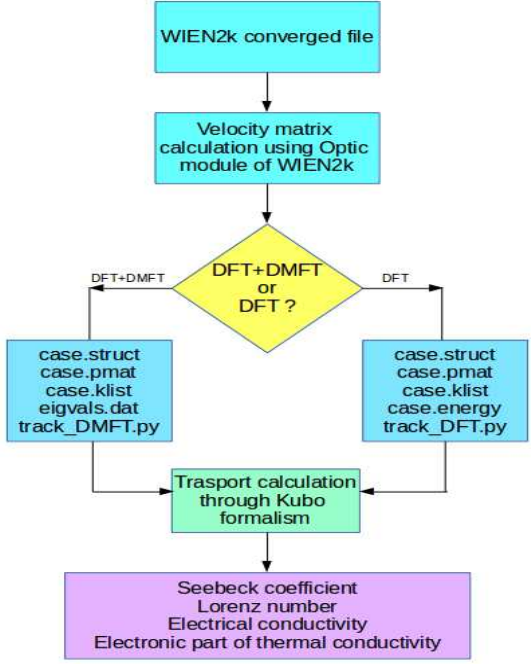} 
    \caption{\small {The workflow of the TRACK code.}}
    \label{fig:}
  \end{center}
\end{figure}

\begin{table}
\caption{\small{The details of different input parameters.}}
\resizebox{0.48\textwidth}{!}{%
\begin{tabular}{@{\extracolsep{\fill}}c c c c c c c c c c cc c c c c c } 
\hline\hline
 
\multicolumn{1}{c}{Name} & & \multicolumn{1}{c}{Unit} & & \multicolumn{1}{c}{Meaning} \\
case & & - & & Case file name \\
mu\_value & & eV & & Zero point of the energy \\
cal\_direction & & - & & Direction: 0$\rightarrow$ $xx$; 1$\rightarrow$ $yy$; 2$\rightarrow$ $zz$ \\
cal\_type & & - & & Type of calculation: 1$\rightarrow$ DFT+DMFT; 2$\rightarrow$DFT \\
beta\_T & & eV & & Inverse of the temperature \\
mag\_order & & - & & 1$\rightarrow$ non-magnetic; 2$\rightarrow$ magnetic \\
spin & & - & & 0$\rightarrow$ non-magnetic; 1$\rightarrow$ spin up; 2$\rightarrow$ spin down\\
omega\_min & & eV & & Minimum energy value \\
omega\_max & & eV & & Maximum energy value \\
omega\_iter & & - & & Division factor for step size of energy value\\
tau\_type & & - & & 1$\rightarrow$constant relaxation time; 2$\rightarrow$ relaxation time\\
tau & & s & & constant value of $\tau$ \\
omega\_pwr & & - & & Power factor for $\omega$ dependent function \\ 
 & & & & related with $Im\, \Sigma (\omega)$\\
omega\_cnst & & - & & Proportionality constant for $Im\, \Sigma (\omega)$ \\
 & &  & & and $\omega$ dependent function \\  
dft\_im\_selfen & & eV & & Imaginary part of self-energy \\
start\_temp & & K & & Value of the starting temperature \\
end\_temp & & K & & Value of the end temperature \\
temp\_step & & - & & Step size for the temperature difference\\
trans\_cal & & - & & F$\rightarrow$ Fresh start \\
 & &  & & R$\rightarrow$ Restart after formatting the energy file \\

\hline

\hline\hline
 
\end{tabular}}
\end{table} 

\section{Theoretical background}

\subsection {Using many-body electronic structure theory}

Here, the Kubo linear-response formalism is utilized for computing different transport properties without considering the vertex correction \cite{tomczak,oudovenko}. Under this formalism without vertex correction, $\sigma$, $S$ and $\kappa_e$ along $\alpha\alpha^\prime$ direction are defined as, 

\begin{equation} 
\sigma^{\alpha \alpha^{\prime}} = \beta e^2K_0^{\alpha \alpha^{\prime}}
\end{equation}

\begin{equation}
S^{\alpha \alpha^{\prime}} = -\frac{k_B}{|e|}\frac{K_1^{\alpha \alpha^{\prime}}}{K_0^{\alpha \alpha^{\prime}}}
\end{equation}

\begin{equation}
\kappa^{\alpha \alpha^{\prime}}_e = k_B\Bigg ( K_2^{\alpha \alpha^{\prime}} - \frac{(K_1^{\alpha \alpha^{\prime}})^2}{K_0^{\alpha \alpha^{\prime}}} \Bigg )
\end{equation}

 where, $\alpha(= x, y, z)$ is the direction, $\beta$ denotes the inverse of temperature (T), $e$ represents the electronic charge and $K_0^{\alpha \alpha^{\prime}}$ ($K_1^{\alpha \alpha^{\prime}}$) shows current-current (current-heat) correlation function. In general, the correlation function is called as kinetic coefficient which is given by,  

\begin{equation}
K_n^{\alpha \alpha^{\prime}} = N_{sp}\pi \hbar \int d\omega (\beta\omega)^nf(\omega)f(-\omega)\mathit\Gamma^{\alpha \alpha^{\prime}}(\omega,\omega)
\end{equation}

  where, $N_{sp}$ and $f(\omega)$ are the spin factor and Fermi function, $\hbar$ denotes the reduced Planck's constant. Along with these, 
$\mathit\Gamma^{\alpha \alpha^{\prime}}$ represents the transport distribution. In matrix notation after neglecting the vertex correction, this quantity is expressed by, 

\begin{equation}
\mathit\Gamma^{\alpha \alpha^{\prime}}(\omega_1,\omega_2) = \frac{1}{V}{\sum\limits_{k} }Tr \bigg (v^\alpha(\mathbf{k})A(\mathbf{k},\omega_1)v^{\alpha^{\prime}}(\mathbf{k})A(\mathbf{k},\omega_2)  \bigg)
\end{equation}

  where $V$ is the unit-cell volume, $v^\alpha(\mathbf{k})$ is the group velocity and $A(\mathbf{k},\omega)$ represents the spectral function. Here, the momentum matrix method is used for finding the $v^\alpha(\mathbf{k})$, which is given by \cite{optic}, 
  
\begin{equation}
\mathbf{\upsilon }_{n}^\alpha(\textbf{k}) = \frac{1}{m_e}\langle\varPsi_{n\textbf{k}}|\hat{\textbf{p}}^\alpha|\varPsi_{n\textbf{k}}  \rangle 
\end{equation}

  where, $m_e$ and $\hat{\textbf{p}}^\alpha$ are mass of free electron and momentum operator along $\alpha$ direction, respectively. $\varPsi_{n\textbf{k}}$ denotes the Kohn-Sham wave function, where n (\textbf{k}) is band index (crystal momentum in reciprocal space). This method is chosen because it avoids the well known \textquotedblleft band-crossing problem\textquotedblright\, as mentioned in the earlier literatures \cite{scheidemantel,transopt}. The similar implementation is carried out in WIEN2k optic module for computing this quantity \cite{optic}. In addition to this, the spectral function in real $\omega$ is expressed by \cite{imada_rev},   

\begin{equation}
A{(\mathbf{k},\omega)} = -{\Huge  \frac{1}{\pi} \frac{Im\, \Sigma (\omega)}{[\omega - \varepsilon^0_\mathbf{k} - Re\, \Sigma (\omega)]^2 + [Im\, \Sigma (\omega)]^2}}
\end{equation}

  where, $\varepsilon^0_\mathbf{k}$ is the energy of non-interacting single electronic state with crystal momentum \textbf{k}, $Im\, \Sigma (\omega)$ \big($Re\, \Sigma (\omega)$\big) represents the imaginary (real) part of self-energy ($\Sigma (\omega)$) obtained from DFT+DMFT calculation. Furthermore, the Lorenz number ($L$) is defined as,

\begin{equation}
L=\frac{\kappa^{\alpha \alpha^{\prime}}_e}{T \sigma^{\alpha \alpha^{\prime}}}
\end{equation}

\begin{table*}
\caption{\small {Calculated values of Seebeck coefficients ($S$) for Vanadium (V) at different temperatures by using TRACK$_{DMFT}$, TRACK$_{DFT}$ and BoltzTraP along with experimental data. (Here, $S$ is in $\mu V/K$).}}
\resizebox{0.65\textwidth}{!}{%
\begin{tabular}{@{\extracolsep{\fill}}c c c c c c c c c c c c c c cc c c c c c c c c } 
\hline\hline
 
\multicolumn{1}{c}{Temperature} & & \multicolumn{1}{c}{Experimental} & & \multicolumn{1}{c}{} & & \multicolumn{1}{c}{Calculated} & & \multicolumn{1}{c}{} \\
\multicolumn{1}{c}{(K)} & & \multicolumn{1}{c}{} & & \multicolumn{1}{c}{TRACK$_{DMFT}$} & & \multicolumn{1}{c}{TRACK$_{DFT}$} & & \multicolumn{1}{c}{BoltzTraP} \\
                             
\hline
200 & & 0.41\cite{jung}, 0.27\cite{okram}, 0.86\cite{mackintosh}, 0.60\cite{mackintosh} & & 6.28 & & 2.50 & & 1.68 \\
250 & & -0.11\cite{jung}, -0.39\cite{okram}, 0.6\cite{mackintosh}, 0.39\cite{mackintosh} & & 7.62 & & 4.34 & & 2.10 \\
288 & & -0.36\cite{jung}, -0.89\cite{okram},0.83\cite{mackintosh},0.55\cite{mackintosh} & & - & & - & & - \\
300 & & -1.06\cite{okram}, 1.0\cite{mackintosh} & & 5.95 & & 5.86 & & 2.49 \\
350 & & - & & 2.65 & & 7.13 & & 2.83 \\
400 & & - & & 1.84 & & 8.23 & & 3.14 \\
450 & & - & & -3.59 & & 9.22 & & 3.38 \\
\hline\hline
 
\end{tabular}}
\end{table*}

\subsection {Using band-theory}

  In order to use the eigenvalues obtained from band-theory's calculation, the transport distribution ($i.e.$ equation 5) of Kubo formalism needs to be modified accordingly. The $v^\alpha(\mathbf{k})$ is already obtained by using single-particle Kohn-Sham orbitals. The two $A{(\mathbf{k},\omega)}$ are seen in the equation of transport distribution, where the form of $A{(\mathbf{k},\omega)}$ is usually known as Lorentzian like function of $\omega$ for particular $\varepsilon^0_\mathbf{k}$. For non-interacting theory, the value of $Re\, \Sigma (\omega)$ is equal to zero. Therefore, to transform one of the $A{(\mathbf{k},\omega)}$ from Lorentzian type to Dirac delta function of $\delta (\omega - \varepsilon^0_\mathbf{k})$, typically very small value of $Im\, \Sigma (\omega)$ must be considered. This step is needed because the value of $Im\, \Sigma (\omega)$ represents the half width at the half maxima of the Lorentzian function and taking small value of this quantity will transform it into Dirac delta function. Another $A{(\mathbf{k},\omega)}$ is replaced by $\big(-\frac{1}{Im\, \Sigma (\omega)}\big)$. Further, the $Im\, \Sigma (\omega)$ is function of $\omega$, which denotes the strength of EEI within material. In general, this quantity is equal to $\big(-\frac{\tau(\omega)}{\hbar}\big)$. Therefore, if all the above mentioned changes are implemented in equations (1-7), then the $\sigma$ for a particular band will retrieve the form of $\sigma$ as obtained from semiclassical theory \cite{aschroft_eq}. Thus, with these approximations, the transport coefficients within Kubo formalism can describe transport parameters which are obtained from the semi-classical BTE with RTA by utilizing the single-particle eigenvalues.

  Further, the convergence of $S$ needs to be checked with varying the value of $Im\, \Sigma (\omega)$ to see whether proper $\delta (\omega - \varepsilon^0_\mathbf{k})$ is obtained or not. The present implementation provides the opportunity to study the importance of $\omega$ dependent $\tau$ for calculating the transport properties. However till now, BoltzTraP and BOLTZWANN don't provide the results with considering $\tau(\omega)$. These two codes are most commonly used for studying transport properties of materials. All these procedures are also applicable for computing the transport coefficients from the estimated eigenvalues of the static mean-field DFT+$U$ method, where $U$ is called as Hubbard parameter.
  
\section{Workflow and Technical details}

\subsection{Workflow}

 In order to compute the transport properties using this code, all the input variables need to be provided in single file, which is named as $track.input$. The details of different input parameters are described in Table 1 and the screenshot of $track.input$ files are shown in Figs. 1(a) and 1(b). Along with this, additional files named as $case.struct$, $case.klist$ and $case.pmat$ are necessary to be kept in the same directory of input file for getting the structural information, a list of k-points in the brillouin zone and the information of velocity matrix elements between all bands in the desired energy window for each k-point, respectively. The $case.struct$ and $case.klist$ are two basic files for any WIEN2k calculation. The detailed step to find the velocity matrix elements are explained in the userguide of WIEN2k \cite{wien2k}. In addition to these, the $eigvals.dat$ file is needed for the transport properties calculations using the eigenvalues of DFT+DMFT. The file contains the information of momentum resolved spectral functions as calculated by eDMFTF code. In similar manner, for performing the transport calculations using the eigenvalues of DFT, $case.energy$ file is needed from the self-consistent WIEN2k calculation along with $track.input$, $case.struct$, $case.klist$ and $case.pmat$ files. In case of DFT+$U$ within WIEN2k, $case.energyup$ or (and) $case.energydn$ file is (files are) needed for finding the transport parameters for non-magnetic (magnetic) material. The workflow of the code can be more clearly seen in the flowchart shown in Fig. 2.

\subsection{Technical details}

  TRACK is implemented in modern Python 3 language, which is incompatible with Python 2. The \textit{Numpy} and \textit{Scipy} libraries are extensively utilized for performing the several numerical operations \cite{numpy,scipy}. Along with these, the \textit{Time} module is used for estimating computation's time. In the current version of the code, it is only interfaced with the WIEN2k and eDMFTF codes.  
  
  Two different modes of calculations are implemented in TRACK code ($i.e.$ TRACK$_{DMFT}$ and TRACK$_{DFT}$), which depends on the choice of eigenvalues from the electronic structure methods. In such scenario, when one uses the eigenvalues of eDMFTF (WIEN2k) code then $track\_DMFT.py$ ($track\_DFT.py$) code needs to be run by the command - $python3$ $track\_DMFT.py$ ($track\_DFT.py$). Dense k-mesh is strongly advised for obtaining the eigenvalues from the electronic structure calculation with checking the convergence of transport coefficients. Proper sampling of $\omega$ around the $E_F$ is needed for transport calculation, because transport behaviours of any material highly depend on low-energy window. It is important to note that when one uses the $track\_DFT.py$ code then the difference between two successive $\omega$ must be equal to $\sim\frac{1}{10}$ times of $Im\, \Sigma (\omega)$. The easy implementation makes this code more user friendly to compute different transport properties.

\section{Test cases}

  Here, we describe the capabilities of the TRACK code by performing calculations on three different types of correlated materials: (i) Vanadium (V), a well known correlated metal; (ii) FeSi, a famous semiconductor showing many exotic phenomena; (iii) LaCoO$_3$, a well-studied correlated system along with metal-insulator transition and good TE properties.
  
  It is known that the values of $\sigma$ and $\kappa_e$ vary based on the estimation of $\tau$. Here, in case of DFT+DMFT methodology, only EEI is considered for getting temperature dependent electronic structure properties of any material. Therefore, in present scenario, the other sources of electronic interactions are not considered for finding the value of temperature dependent $\tau$, which may lead to the bad prediction of these two quantities from the TRACK$_{DMFT}$ code. Thus, we calculate temperature dependent $L$ parameter using this code, which is a independent quantity of $\tau$ value like $S$.  
\begin{table*}
\caption{\small {Calculated values of Seebeck coefficients ($S$) for FeSi at different temperatures by using TRACK$_{DMFT}$, TRACK$_{DFT}$ and BoltzTraP along with experimental data. (Here, $S$ is in $\mu V/K$).}}
\resizebox{0.70\textwidth}{!}{%
\begin{tabular}{@{\extracolsep{\fill}}c c c c c c c c c c c c c c cc c c c c c c c c } 
\hline\hline
 
\multicolumn{1}{c}{Temperature} & & \multicolumn{1}{c}{Experimental} & & \multicolumn{1}{c}{} & & \multicolumn{1}{c}{Calculated} & & \multicolumn{1}{c}{} \\
%\multicolumn{1}{c}{} & & \multicolumn{1}{c}{Seebeck Coffecient} & & \multicolumn{1}{c}{} & & \multicolumn{1}{c}{Seebeck Coffecient} & & \multicolumn{1}{c}{} \\
\multicolumn{1}{c}{(K)} & & \multicolumn{1}{c}{} & & \multicolumn{1}{c}{TRACK$_{DMFT}$} & & \multicolumn{1}{c}{TRACK$_{DFT}$} & & \multicolumn{1}{c}{BoltzTraP} \\
                             
\hline
100 & & 25.0\cite{buschinger_fesi}, 41.3\cite{yang_fesi}, 23.6\cite{wolfe_fesi}, 20.2\cite{tomczak_fesi} & & - & & 13.5 & & -31.2 \\
110 & & -1.8\cite{buschinger_fesi}, -2.6\cite{yang_fesi}, 40.0\cite{sales_fesi}, 8.8\cite{wolfe_fesi}, 11.8\cite{tomczak_fesi} & & 11.7 & & - & & - \\
180 & & -12.4\cite{buschinger_fesi}, -17.1\cite{yang_fesi}, -13.3\cite{sales_fesi}, -8.1\cite{wolfe_fesi}, -6.6\cite{tomczak_fesi} & & 3.6 & & - & & - \\
200 & & -10.0\cite{buschinger_fesi}, -12.6\cite{yang_fesi}, -10.4\cite{sales_fesi}, -8.8\cite{wolfe_fesi}, -7.0\cite{tomczak_fesi} & & - & & 2.6 & & -49.1 \\
290 & & 2.1\cite{buschinger_fesi}, 0.6\cite{yang_fesi}, 3.4\cite{wolfe_fesi}, -6.2\cite{tomczak_fesi} & & -33.5 & & - & & - \\
300 & & 0.1\cite{yang_fesi}, -2.1\cite{sales_fesi}, 3.5\cite{wolfe_fesi}, -6.5\cite{tomczak_fesi} & & -23.2 & & -5.6 & & -35.0 \\
380 & & 2.9\cite{sales_fesi}, -13.1\cite{tomczak_fesi} & & 10.9 & & - & & - \\
400 & & 3.6\cite{sales_fesi} & & - & & -8.6 & & -24.4 \\
600 & & 11.3\cite{sales_fesi}, 0.5\cite{tomczak_fesi} & & 11.3 & & -9.2 & & -12.7 \\
700 & & 14.7\cite{sales_fesi}, 2.4\cite{tomczak_fesi} & & 6.0 & & -8.3 & & -8.8 \\
800 & & 4.96\cite{tomczak_fesi} & & 18.5 & & -6.9 & & -5.2 \\
1000 & & 12.5\cite{tomczak_fesi} & & 30.3 & & -3.8 & & 2.0 \\

\hline\hline
 
\end{tabular}}
\end{table*}

\begin{table*}
\caption{\small {Calculated values of Seebeck coefficients ($S$) for LaCoO$_3$ at different temperatures by using TRACK$_{DMFT}$, TRACK$_{DFT}$ and BoltzTraP along with experimental data. (Here, $S$ is in $\mu V/K$).}}
\resizebox{0.72\textwidth}{!}{%
\begin{tabular}{@{\extracolsep{\fill}}c c c c c c c c c c c c c c cc c c c c c c c c } 
\hline\hline
 
\multicolumn{1}{c}{Temperature} & & \multicolumn{1}{c}{Experimental} & & \multicolumn{1}{c}{} & & \multicolumn{1}{c}{Calculated} & & \multicolumn{1}{c}{} \\
\multicolumn{1}{c}{(K)} & & \multicolumn{1}{c}{} & & \multicolumn{1}{c}{TRACK$_{DMFT}$} & & \multicolumn{1}{c}{TRACK$_{DFT}$} & & \multicolumn{1}{c}{BoltzTraP} \\
                             
\hline
150 & & 207.9\cite{cortes_lco} & & - & & 15.78 & & -413.25 \\ 
180 & & 431.9\cite{cortes_lco} & & -165.6 & & - & & - \\
200 & & 558.3\cite{cortes_lco} & & - & & 24.65 & & -644.41 \\
250 & & 623.3\cite{cortes_lco} & & - & & 58.53 & & -579.46 \\
290 & & 578.5\cite{cortes_lco} & & 93.5 & & - & & - \\
300 & & -351.1\cite{he_lco}, -323.7\cite{ohtani_lco}, 249.3\cite{singh_mes_lco}, 524.8\cite{cortes_lco}, 615.6\cite{sk_high_lco} & & 138.0 & & 159.8 & & -529.15 \\
350 & & -230\cite{fu_lco}, -219.8\cite{he_lco}, -69.5\cite{ohtani_lco}, 287.1\cite{singh_mes_lco}, 474.1\cite{sk_high_lco} & & -  & & 309.6 & & -482.0 \\
400 & & -121.5\cite{fu_lco}, -15.96\cite{he_lco}, 111.1\cite{ohtani_lco}, 218.2\cite{singh_mes_lco}, 275.3\cite{sk_high_lco} & & 241.14 & & 418.45 & & -438.35  \\
450 & & -3.6\cite{fu_lco}, 122.7\cite{he_lco}, 156.8\cite{ohtani_lco}, 121.5\cite{singh_mes_lco}, 147.5\cite{sk_high_lco} & & - & & 459.04 & & -398.20 \\
500 & & 54.9\cite{fu_lco}, 177.3\cite{he_lco}, 70.5\cite{singh_mes_lco}, 83.9\cite{sk_high_lco} & & 294.6 & & 457.81 & & -361.52  \\
550 & & 60.5\cite{fu_lco}, 159.5\cite{he_lco}, 49.7\cite{singh_mes_lco}, 54.7\cite{sk_high_lco} & & - & & 438.64 & & -328.31 \\
600 & & 52.9\cite{fu_lco}, 132.6\cite{he_lco}, 34.7\cite{singh_mes_lco}, 41.7\cite{sk_high_lco} & & 258.15 & & 413.53 & & -298.49 \\
650 & & 45.2\cite{fu_lco}, 33.8\cite{sk_high_lco} & & - & & 387.60 & & -271.93 \\
700 & & 26.2\cite{sk_high_lco} & & 131.62 & & 362.87 & & -248.40 \\
750 & & 20.2\cite{sk_high_lco} & & - & & 340.03 & & -227.65 \\
800 & & 14.9\cite{sk_high_lco} & & 121.98 & & 319.24 & & -209.42 \\

\hline\hline
 
\end{tabular}}
\end{table*} 

\begin{figure*}
  \begin{center}
    \includegraphics[width=0.85\linewidth, height=4.8cm]{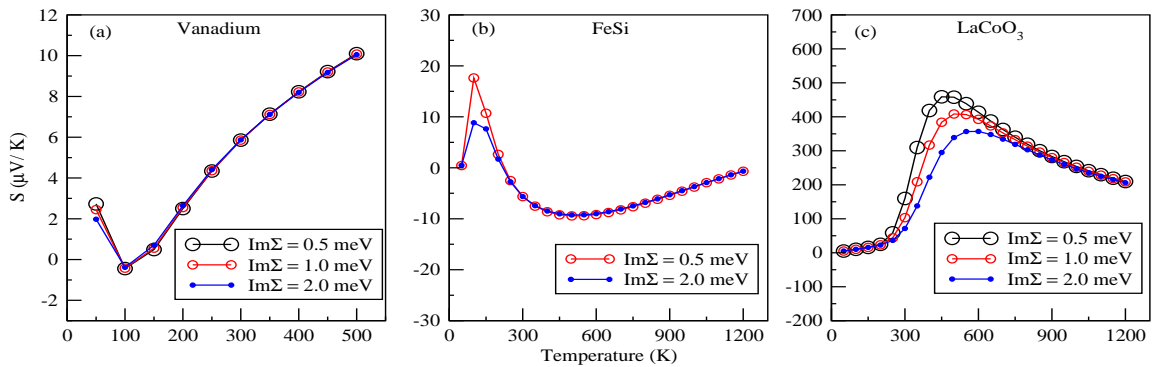} 
    \caption{\small {Temperature dependent Seebeck coefficient ($S$) of (a) Vanadium (V) (b) FeSi and (c) LaCoO$_3$ with different values of $Im\, \Sigma (\omega)$ for checking the convergence.}}
    \label{fig:}
  \end{center}
\end{figure*}

\begin{table}
\caption{\small {Calculated values of temperature dependent Lorenz numbers ($L$) by using TRACK$_{DMFT}$, TRACK$_{DFT}$ and BoltzTraP for Vanadium. (Here, $L$ is in $\times$10$^{-8}$ $V^2/K^2$ )}}
\resizebox{0.40\textwidth}{!}{%
\begin{tabular}{@{\extracolsep{\fill}}c c c c c c c c c c cc c c c c c c c c } 
\hline\hline
 
\multicolumn{1}{c}{Temperature} & & \multicolumn{1}{c}{} & & \multicolumn{1}{c}{Calculated} & & \multicolumn{1}{c}{} \\
\multicolumn{1}{c}{(K)} & & \multicolumn{1}{c}{TRACK$_{DMFT}$} & & \multicolumn{1}{c}{TRACK$_{DFT}$} & & \multicolumn{1}{c}{BoltzTraP} \\
                             
\hline
200 & & 2.4 & & 2.472 & & 2.43 \\
250 & & 2.3 & & 2.466 & & 2.43 \\
300 & & 2.2 & & 2.456 & & 2.42 \\
350 & & 2.15 & & 2.448 & & 2.40 \\
400 & & 2.1 & & 2.438 & & 2.39 \\
450 & & 2.0 & & 2.425 & & 2.37 \\
\hline\hline
 
\end{tabular}}
\end{table}

\subsection{Computational Details}

  The detailed structural informations of V, FeSi and LaCoO$_3$ are taken from Refs. 26, 27 and 28, respectively for carrying out the present calculations. 92$\times$92$\times$92, 74$\times$74$\times$74 and 53$\times$53$\times$53 k-mesh sizes are used for calculating the transport properties of V, FeSi and LaCoO$_3$, respectively. For DFT+DMFT calculation, the values of $U$=3.4 ($J$=0.52) eV, $U$=4.4 ($J$=0.89) eV and $U$=6.9 ($J$=0.3) eV are set for V 3$d$, Fe 3$d$ and Co 3$d$ electrons of V, FeSi and LaCoO$_3$, respectively. It is noted that the semiconducting ground state of FeSi is found from the DFT calculation, whereas DFT+DMFT provides the semi-metallic behaviour from the temperature dependent study\cite{dutta_fesi}. Here, $E_F$ is fixed at middle of the bandgap for all calculations where bandgap is clearly observed from total density of states.  
  
  In case of LaCoO$_3$, the presence of octahedral rotation in the compound brings important task to decrease the strength of crystal-field splitting\cite{chakra_prm}. Therefore, in order to overcome this with observing the metal-insulator transition, the value of $J$=0.3 eV is chosen for DFT+DMFT. Moreover, it is seen that the experimental data of transport properties are nicely described for this value of $J$ with $U$=6.9 eV, which is the another reason to take these values for the present study. To perform DFT+$U$ calculation for LaCoO$_3$, we followed the work of Singh \textit{et al.}\cite{singh_lco_phlio}. The eigenvalues of this calculation are used for TRACK$_{DFT}$ and BoltzTraP codes to estimate the transport parameters. 
  
  In order to compare with the computed values from the BoltzTraP, the constant value of $\tau$=$10^{-14}$ s is considered for computing the transport properties, when eigenvalues of DFT or DFT+$U$ are used. The convergence of transport properties with varying the value of $Im\, \Sigma (\omega)$ is checked and also advised for the users. All test cases are performed by considering the direction $\alpha = x$ and $\alpha^\prime = x$. Here, all DFT+DMFT calculations are performed using density-density type of Coulomb interaction within eDMFTF code. To get $A{(\mathbf{k},\omega)}$ in real axis, maximum entropy method is used for analytical continuation \cite{jarrell_maxent}.

\subsection {Seebeck Coefficient ($S$)}

  The computed values of temperature dependent $S$ for V are tabulated in Table. II along with the experimental data for comparison. The experimental values of $S$ show decreasing behaviour with increasing temperature for high purity sample, where the sign of this physical property changes at $\sim$250 K \cite{jung,okram}. However, for polycrystalline sample, the values of this quantity is first decreasing till $\sim$250 K and then increasing with rise in temperature \cite{mackintosh,vedernikov}. Moreover, the computed values of $S$ for different temperatures using TRACK$_{DMFT}$ decreases with rise in temperature as similar to experimental data, where the change in the sign of this quantity is found at $\sim$450 K. The estimated value of $S$ is $\sim$6.3 ($\sim$-3.6) $\mu$V/K at $\sim$200 ($\sim$450) K, where the experimental values are $\sim$0.5 ($\sim$-0.36) $\mu$V/K at $\sim$200 ($\sim$289) K. 
  
  For FeSi, Table III represents the calculated values of temperature dependent $S$ with corresponding experimentally reported data. From the table, the different experimental and theoretical results are showing the change of sign of $S$ value in between 110-180 K (from positive to negative) and 290-380 K (from negative to positive), respectively. This behaviour is nicely captured by using the TRACK$_{DMFT}$ as evident from the table. The computed values of $S$ at 300 K using this code is found to be $\sim$-23.2 $\mu$V/K, where the experimental range of $S$ at this temperature is obtained from $\sim$-2.1 to $\sim$3.5 $\mu$V/K. Therefore, the overall trend of computed value using TRACK$_{DMFT}$ is in good agreement with the experimental data for the studied temperature range.
  
  Numerous experimental works on the transport properties of LaCoO$_3$ have been reported. The huge range of $S$ value reveals from the different experimental works for 300-600 K. It happens for employing different sintering temperature to make the sample. The calculated values of $S$ as function of temperature along with experimental data are provided in Table IV. In case of high sintering temperature ($i.e.$ $\geq$1100 K), it is observed that the $S$ values show increasing behaviour with rise in temperature till $\sim$500 K and after this temperature, the decreasing trend in $S$ is found for the studied temperature range \cite{fu_lco, he_lco, ohtani_lco, singh_mes_lco}. The similar behaviour is seen for polycrystalline sample of this compound, where the above mentioned change in $S$ is found after 250 K\cite{cortes_lco}. Here, the TRACK$_{DMFT}$ provides the computed value of $S$ = $\sim$-165.6 ($\sim$130) $\mu$V/K at 180 K (300 K). Moreover, the decreasing nature for $S$ value is obtained after 500 K. At 500 K, it is estimated to be $\sim$294.6 $\mu$V/K, where the experimental range is found to be $\sim$54.9 to $\sim$177.3 $\mu$V/K. Therefore, it is found that the overall temperature dependent behaviour of $S$ is nicely explained using the estimated values of TRACK$_{DMFT}$. 
  
  At this point, it is noted that by taking the appropriate values of $U$, $J$ and changing the type of Coulomb interaction within DFT+DMFT, the predicted values of TRACK$_{DMFT}$ for different class of materials may be tuned to the values more closer to the experimentally reported value. The presence of off-stoichiometry and impurity in the samples also affects the transport properties. Further, it is mentioned that the estimation of bandgap using DFT+DMFT method needs to get proper value for the material, which may be the reason for getting deviation within the calculated and experimental data of the transport parameter.
 
  For TRACK$_{DFT}$, the convergence of $S$ as function of temperature is checked for different values of $Im\, \Sigma (\omega)$ for all three materials, which are shown in Fig. 3(a), 3(b) and 3(c) for V, FeSi and LaCoO$_3$, respectively. The figures show for V and FeSi that $Im\, \Sigma (\omega)$= 0.5 meV provides good convergence of the $S$ for studied temperature range. But, for LaCoO$_3$, the convergence is obtained, when the value of $Im\, \Sigma (\omega)$ is reduced from 2.0 to 0.5 meV, for all the studied temperature except for 300-500 K. Moreover, it is expected that the further reduction of $Im\, \Sigma (\omega)$ will provide good convergence for the mentioned temperature range. But, due to our present limitations of computational resources, we describe the computed values of TRACK$_{DFT}$ code with the fixed value of $Im\, \Sigma (\omega)$=0.5 meV for all the materials.   

\begin{figure*}
  \begin{center}
    \includegraphics[width=0.85\linewidth, height=4.8cm]{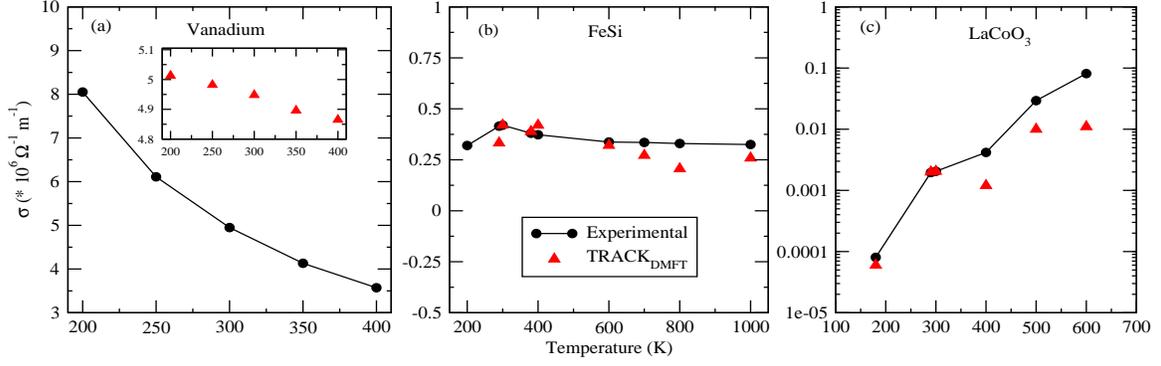} 
    \caption{\small {Calculated values of temperature dependent electrical conductivity ($\sigma$) by using TRACK$_{DMFT}$ for (a) Vanadium (V) (b) FeSi and (c) LaCoO$_3$ along with corresponding experimental data \cite{v_resistivity,wolfe_fesi,lco_resistivity}.}}
    \label{fig:}
  \end{center}
\end{figure*}

\begin{table}
\caption{\small {Calculated values of temperature dependent Lorenz numbers ($L$) by using TRACK$_{DMFT}$, TRACK$_{DFT}$ and BoltzTraP for FeSi. (Here, $L$ is in $\times$10$^{-8}$ $V^2/K^2$)}}
\resizebox{0.40\textwidth}{!}{%
\begin{tabular}{@{\extracolsep{\fill}}c c c c c c c c c c cc c c c c c c c c } 
\hline\hline
 
\multicolumn{1}{c}{Temperature} & & \multicolumn{1}{c}{} & & \multicolumn{1}{c}{Calculated} & & \multicolumn{1}{c}{} \\
%\multicolumn{1}{c}{} & & \multicolumn{1}{c}{Seebeck Coffecient} & & \multicolumn{1}{c}{} & & \multicolumn{1}{c}{Seebeck Coffecient} & & \multicolumn{1}{c}{} \\
\multicolumn{1}{c}{(K)} & & \multicolumn{1}{c}{TRACK$_{DMFT}$} & & \multicolumn{1}{c}{TRACK$_{DFT}$} & & \multicolumn{1}{c}{BoltzTraP} \\
                             
\hline
100 & & 4.35 & & 15.3 & & 18.9 \\
110 & & 6.52 & & - & & - \\
180 & & 3.83 & & - & & - \\
200 & & 2.59 & & 8.8 & & 10.1\\
290 & & 2.71 & & - & & - \\
300 & & 3.72 & & 5.6 & & 7.6 \\
380 & & 3.27 & & - & & -\\
400 & & 2.51 & & 4.2 & & 6.4 \\
600 & & 1.63 & & 3.3 & & 5.1 \\
700 & & 1.67 & & 3.1 & & 4.7 \\
800 & & 2.17 & & 3.0 & & 4.4 \\
1000 & & 1.28 & & 2.7 & & 3.8\\

\hline\hline
 
\end{tabular}}
\end{table}

\begin{figure*}
  \begin{center}
    \includegraphics[width=0.85\linewidth, height=4.8cm]{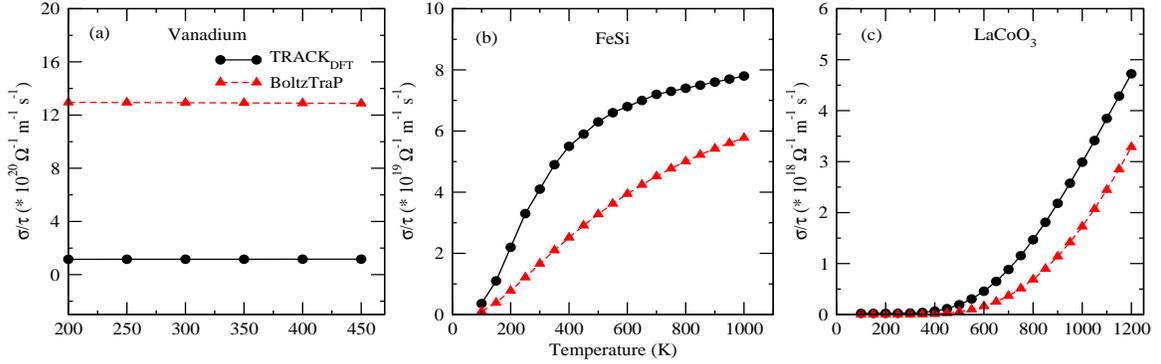} 
    \caption{\small {Calculated values of temperature dependent electrical conductivity per relaxation time ($\sigma/\tau$) by using TRACK$_{DFT}$ and BoltzTraP for (a) Vanadium (V) (b) FeSi and (c) LaCoO$_3$.}}
    \label{fig:}
  \end{center}
\end{figure*} 

  The computed values of $S$ using TRACK$_{DFT}$ code for V are tabulated in Table. II along with the calculated values from BoltzTraP. These values are obtained for the temperature range of 200-450 K. The table shows higher values for TRACK$_{DFT}$ than the BoltzTraP for observed temperature window. The computed values from both these codes represent the increasing trend with rise in temperature, which are providing opposite behaviour than that of experimental data for pure sample. 
  
  Similarly, Table III contains the calculated values of this transport parameter using TRACK$_{DFT}$ for FeSi. In lower temperature region ($i.e.$ $\lesssim$ 300 K), the values are closely matched with the experimental data. But, above this temperature, experimentally it is seen that the sign of $S$ is changed from negative to positive, where the calculated values from TRACK$_{DFT}$ are showing negative sign for 400-1000 K. However, the values are monotonically increasing with rise in temperature, which shows similar fashion as observed from the experimental data for temperature above 300 K. Moreover, BoltzTraP gives negative sign of $S$ for 100-800 K, which does not follow the experimental data. But, the increasing behaviour in $S$ is observed after 200 to 1000 K as similar to experiment. The trend of the $S$ value for TRACK$_{DFT}$ shows quite good matching with the BoltzTraP as well as with the experimental data of this semiconductor. 
    
  In case of LaCoO$_3$, the estimated values of $S$ as function of temperature using TRACK$_{DFT}$ are listed in Table IV. The table represents that the trend wise good matching is seen in between the calculated values of TRACK$_{DFT}$ and experimentally reported data. The value of $S$ at 300 K is found to be $\sim$309.6 $\mu$V/K, where the experimental range at this temperature is found to be $\sim$-230 to $\sim$474.1 $\mu$V/K. However, at the same value of $E_F$, BoltzTraP fails to provide good description of the experimental data as evident from the table. The decreasing nature is seen in the computed values of TRACK$_{DFT}$ above 450 K, where this behaviour is nicely following the experimental data. But, the estimated values of $S$ above 450 K are sufficiently larger than experimental result.
  
  Here, the transport properties of different class of materials with various range of $S$ values are studied using TRACK$_{DMFT}$ and TRACK$_{DFT}$ codes. The above discussion reveals that the trend wise nicely matching is observed with experimental data from the calculated values of TRACK$_{DMFT}$ for all these materials. Moreover, the comparative studies suggest the importance of electronic correlation for studying the $S$ of the correlated materials.

\begin{figure*}
  \begin{center}
    \includegraphics[width=0.85\linewidth, height=5.0cm]{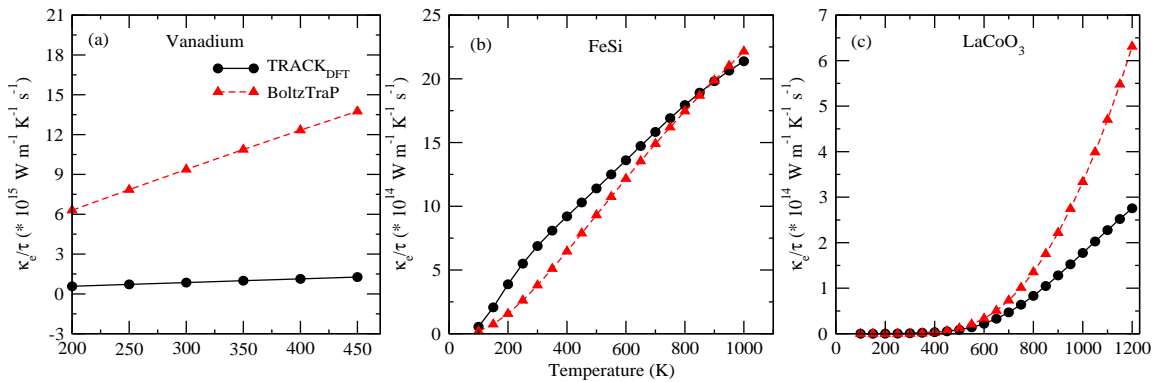} 
    \caption{\small {Calculated values of temperature dependent electronic part of thermal conductivity per relaxation time ($\kappa_e/\tau$) by using TRACK$_{DFT}$ and BoltzTraP for (a) Vanadium (V) (b) FeSi and (c) LaCoO$_3$.}}
    \label{fig:}
  \end{center}
\end{figure*}

\begin{table}
\caption{\small {Calculated values of temperature dependent Lorenz numbers ($L$) by using TRACK$_{DMFT}$, TRACK$_{DFT}$ and BoltzTraP for Vanadium. (Here, $L$ is in $\times$10$^{-8}$ $V^2/K^2$)}}
\resizebox{0.45\textwidth}{!}{%
\begin{tabular}{@{\extracolsep{\fill}}c c c c c c c c c c cc c c c c c c c c } 
\hline\hline
 
\multicolumn{1}{c}{Temperature} & & \multicolumn{1}{c}{} & & \multicolumn{1}{c}{Calculated} & & \multicolumn{1}{c}{} \\
%\multicolumn{1}{c}{} & & \multicolumn{1}{c}{Seebeck Coffecient} & & \multicolumn{1}{c}{} & & \multicolumn{1}{c}{Seebeck Coffecient} & & \multicolumn{1}{c}{} \\
\multicolumn{1}{c}{(K)} & & \multicolumn{1}{c}{TRACK$_{DMFT}$} & & \multicolumn{1}{c}{TRACK$_{DFT}$} & & \multicolumn{1}{c}{BoltzTraP} \\
                             
\hline
180 & & 26.2 & & - & & - \\
200 & & - & & 2.96 & & 159 \\
290 & & 15.02 & & - & & - \\
300 & & 14.36 & & 11.3 & & 84.8 \\
400 & & 10.1 & & 12.9 & & 56.5 \\
500 & & 2.4 & & 8.95 & & 42.4 \\
600 & & 2.9 & & 8.0 & & 34.0 \\
700 & & 4.4 & & 7.6 & & 28.5 \\
800 & & 1.0 & & 7.1 & & 24.5 \\

\hline\hline
 
\end{tabular}}
\end{table}

\subsection{Electrical conductivity ($\sigma$)} 

  The experimental values of $\sigma$ for V, FeSi and LaCoO$_3$ are shown in Figs. 4(a), 4(b) and 4(c), respectively, along with the computed values of TRACK$_{DMFT}$. The code provides large value of $\sigma$ due to only considering the EEI. But, whether the trend of the temperature dependent experimental $\sigma$ depends on the EEI or not, the computed value of $\sigma$ at 300 K is divided by experimental $\sigma$ of similar temperature. In addition to this, the factor obtained by this operation is used for rescaling the computed values of all studied temperatures. The factor is found to be 296.7, 2883.6, 282146.3 for V, FeSi and LaCoO$_3$, respectively. The figures corresponding to these three materials represent that the shape of the temperature dependent experimental $\sigma$ is nicely described by the computed values. Therefore, the results suggest the importance of EEI for providing the trend of the $\sigma$ for all the sample materials.

  The $\sigma/\tau$ as function of temperatures, which are calculated by using TRACK$_{DFT}$ and BoltzTraP, are shown in Fig. 5 for three materials. In case of V, BoltzTraP code gives higher values of $\sigma/\tau$ than the TRACK$_{DFT}$ code, which is also found from Fig 5(a). The computed values of $\sigma/\tau$ using TRACK$_{DFT}$ (BoltzTraP) for this metal is $\sim$1.16 ($\sim$12.9) \texttimes 10$^{20}$ $\Omega^{-1} m^{-1} s^{-1}$ for the studied temperature window. For FeSi, Fig. 5(b) illustrates the $\sigma/\tau$ as function of temperature, where TRACK$_{DFT}$ estimates larger value than BoltzTraP for the studied temperature range. The temperature dependent curve  of $\sigma/\tau$ for TRACK$_{DFT}$ nicely follows BoltzTraP for this semiconductor. The computed values of temperature dependent $\sigma/\tau$ for LaCoO$_3$ is plotted in Fig. 5(c), where TRACK$_{DFT}$ and BoltzTraP codes provide the increasing nature with rise in temperature. The deviation within the computed values of these two codes is found after 600 K. The TRACK$_{DFT}$ gives higher value than BoltzTraP for studied temperature window. Therefore, it is evident from the above discussion that these transport properties can be explained by the TRACK$_{DFT}$ as similar to BoltzTraP for different types of materials.

\subsection{Electronic part of thermal conductivity ($\kappa_e$) and Lorenz number ($L$)} 

  The calculated values of $\kappa_e/\tau$ by using TRACK$_{DFT}$ and BoltzTraP are shown in Fig. 6 for sample materials. For V, the BoltzTraP provides higher values than TRACK$_{DFT}$ for all studied temperature, which is evident from Fig. 6 (a). But, the increasing behaviour with the rise in temperature is observed from the calculated values. The estimated values by utilizing TRACK$_{DFT}$ (BoltzTraP) are increasing from $\sim$0.57 ($\sim$6.3) to $\sim$1.27 ($\sim$13.7) \texttimes 10$^{15}$ $W m^{-1} K^{-1} s^{-1}$ on changing the temperature from 200 K to 450 K. Therefore for V, the order of this physical parameter is found to be similar for both cases. Fig. 6(b) illustrates the $\kappa_e/\tau$ as function of temperature for FeSi. The figure represents that the calculated values from both the codes are almost same except for 200-600 K, where TRACK$_{DFT}$ is showing slightly higher value than BoltzTraP. Therefore, it is noted that the behaviour of the curve as obtained from TRACK$_{DFT}$ is nicely matching with the curve of BoltzTraP for this transport coefficient. The comparison between the calculated values of temperature dependent $\kappa_e/\tau$ by using TRACK$_{DFT}$ and BoltzTraP codes is plotted in Fig. 6(c) for LaCoO$_3$. These two codes provide the increasing nature with rise in temperature for this physical quantity. The slope of the curve and the values for the studied temperatures of $\kappa_e/\tau$ of TRACK$_{DFT}$ are lower than BoltzTraP. For LaCoO$_3$, the major deviation is found after 600 K for this transport parameter. Therefore, overall features of $\kappa_e/\tau$ curve, which are computed using TRACK$_{DFT}$, for test materials are in good match with the results of BoltzTraP.
  
  Further, the calculated values of temperature dependent $L$ by using TRACK$_{DMFT}$, TRACK$_{DFT}$ and BoltzTraP codes for V, FeSi and LaCoO$_3$ are shown in Table V, Table VI and Table VII, respectively. The Table V shows decreasing behaviour of $L$ with increasing temperature. The computed value of $L$ is found to be $\sim$2.4 ($\sim$2.472) \texttimes 10$^{-8}$ V$^2$/K$^2$ at 200 K from TRACK$_{DMFT}$ (TRACK$_{DFT}$) code, whereas BoltzTraP provides $\sim$2.43 \texttimes 10$^{-8}$ V$^2$/K$^2$ at the same temperature. The calculated values of $L$ for different temperatures obtained from the present code are in good agreement with the experimental data \cite{jung} of V. Next for FeSi, it is seen from Table VI that the values obtained from TRACK$_{DMFT}$ calculation are showing good agreement with the typical order of metals and degenerate semiconductors. The value of $L$ at 300 K is estimated to be $\sim$3.72 \texttimes 10$^{-8}$ V$^2$/K$^2$ from TRACK$_{DMFT}$. Moreover, both TRACK$_{DFT}$ and BoltzTraP codes provide larger value than TRACK$_{DMFT}$ as seen from the Table VI. The computed value is found to be $\sim$5.6 ($\sim$7.6) \texttimes 10$^{-8}$ V$^2$/K$^2$ by using TRACK$_{DFT}$ (BoltzTraP) at 300 K. In case of LaCoO$_3$, the Table VII provides the calculated values of $L$=1.0 - 4.4 \texttimes 10$^{-8}$ V$^2$/K$^2$ for 500 - 800 K using TRACK$_{DMFT}$, which is typical order of this parameter for any metal. Similarly, the TRACK$_{DFT}$ gives slightly higher value than TRACK$_{DMFT}$ for this temperature range, whereas BoltzTraP predicts larger value than the typical value of $L$ for the metal. Moreover, from 180-400 K, the computed value of $L$ using TRACK$_{DMFT}$ is found to decrease from 26.2 \texttimes 10$^{-8}$ V$^2$/K$^2$ to 10.1 \texttimes 10$^{-8}$ V$^2$/K$^2$. This large value of $L$ may be obtained due to the underestimation of the experimental bandgap within this temperature window, where insulating phase is seen in DFT+DMFT calculation. However, BoltzTraP shows nonphysical large values of $L$ for the studied temperature window. Thus, the result suggests the better predictive power of TRACK code for computing the temperature dependent $L$ value.   

\section{Conclusions} 

 We presented a Python 3 code TRACK for calculating the transport properties of different class of materials. Kubo linear-response formalism is used for calculating the transport coefficients. The benefit of using this formalism is to access all types of correlated materials ($i.e.$ weak, moderate and strong) with help of interacting and non-interacting electronic structure methodology. By introducing appropriate approximations in the Kubo formalism the transport expression under relaxation time is obtained. The approximated Kubo formalism provides promising results when band-theory is utilized as \textit{ab-initio} method. Here, Vanadium (V), FeSi and LaCoO$_3$ are chosen as test samples to benchmark the TRACK code. The results obtained from TRACK$_{DFT}$ are also compared with the calculated values of BoltzTraP code. The behaviour of different temperature dependent transport parameters are showing good agreement with the trend of corresponding experimental results for all three materials. Therefore, this code is simultaneously useful to understand the basic physics of transport properties for many emergent materials as well as predicting the values of temperature dependent transport parameters for application purpose. 
 
\section{References}
\small
%\bibliography{ms2}
%\bibliographystyle{ieeetr}

\end{document}